\documentclass{InterspeechUpdated}

\interspeechcameraready

\usepackage[utf8]{inputenc} % allow utf-8 input
\usepackage[T1]{fontenc}    % use 8-bit T1 fonts
\usepackage{hyperref}       % hyperlinks
\usepackage{url}            % simple URL typesetting
\usepackage{booktabs}       % professional-quality tables
\usepackage{amsfonts}       % blackboard math symbols
\usepackage{nicefrac}       % compact symbols for 1/2, etc.
\usepackage{microtype}      % microtypography
\usepackage{xcolor}         % colors
\usepackage{cite} % [1, 2, 3, 4, 5] -> [1-5]
\hypersetup{colorlinks=true}
\usepackage[page]{appendix}
\usepackage{caption}
\usepackage{subcaption}
\usepackage{graphicx}
\usepackage{algorithm}
\usepackage{svg}
\usepackage[noend]{algpseudocode}
\usepackage{fancybox,framed}
\usepackage{multirow}
\usepackage{tabularx}
\usepackage{soul}
\usepackage{amsmath}
\usepackage{enumitem}
\usepackage{multirow}
\usepackage{diagbox}
\usepackage[export]{adjustbox}
\usepackage{array}
\usepackage{colortbl}

\usepackage{float}
\floatstyle{plaintop}
\restylefloat{table}

\usepackage{tipa}
\usepackage[most]{tcolorbox}
\usepackage{makecell}

\title{GraphemeAug: A Systematic Approach to Synthesized Hard Negative Keyword Spotting Examples}
\author[]{Harry}{Zhang}
\author[]{Kurt}{Partridge}
\author[]{Pai}{Zhu}
\author[]{Neng}{Chen}
\author[]{Hyun Jin}{Park}
\author[]{Dhruuv}{Agarwal}
\author[]{Quan}{Wang}

\affiliation[nocounter]{Google DeepMind}{Mountain View, CA}{U.S.A}
\email{\{harryz, kep, paizhu, nengchen, hjpark, dhruuv, quanw\}@google.com}
\keywords{keyword spotting, confusing word, text to speech}

\renewcommand{\textrightarrow}{$\rightarrow$}
\usepackage{comment}

\begin{document}

\maketitle

\begin{abstract}

Spoken Keyword Spotting (KWS) is the task of distinguishing between the presence and absence of a keyword in audio. The accuracy of a KWS model hinges on its ability to correctly classify examples close to the keyword and non-keyword boundary. These boundary examples are often scarce in training data, limiting model performance. In this paper, we propose a method to systematically generate adversarial examples close to the decision boundary by making insertion/deletion/substitution edits on the keyword's graphemes. We evaluate this technique on held-out data for a popular keyword and show that the technique improves AUC on a dataset of synthetic hard negatives by 61\% while maintaining quality on positives and ambient negative audio data.

\end{abstract}
\begin{figure*}[tbp]
\centering
\includegraphics[width=0.9\textwidth]{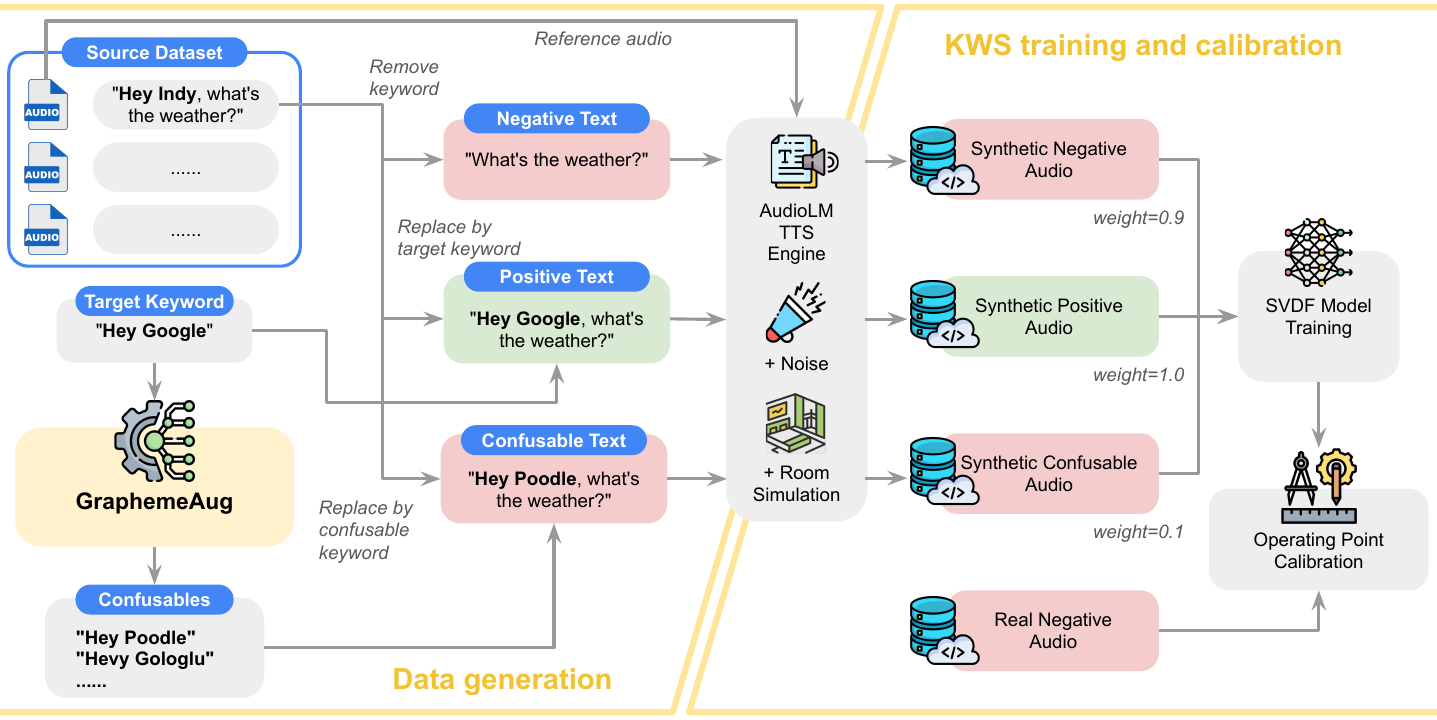}
\caption{The data generation process to create positive, negative, and confusable synthetic datasets. This paper proposes incorporating the GraphemeAug algorithm (lower left corner), which is able to systematically generate a set of confusables.}
\label{fig:data-creation-figure}
\vspace{-10pt}
\end{figure*}

\section{Introduction}
\label{sec:intro}

Keyword spotting is a critical component in voice-activated systems, responsible for accurately detecting predefined keywords, such as ``Alexa'' or ``Siri'' to initiate actions. While KWS systems have achieved significant progress with neural network architectures~\cite{raziel2019endtoendstreamingkeywordspotting, panchapagesan16_interspeech, wang2017smallfootprintkeywordspottingusing,zhu2024ge2e}, they can perform poorly on ``confusable'' phrases---words that sound similar to the keyword and trigger false activations, such as mistaking ``All exhausted...'' for ``Alexa'' or ``Okay, poodle'' for ``Okay, Google''. Past studies have wrestled with the problem of similar sounding phrases by constructing adversarial examples generated from existing examples~\cite{wang2024disentangledtrainingadversarialexamples, 8683479, ai2024mmkwsmultimodalpromptsmultilingual, 9053009}. Recent studies have taken a different approach of creating synthetic audio from a pre-selected list of confusables. Both have been shown effective at reducing false accept rates~\cite{jia2020trainingwakeworddetection, wang2022generatingadversarialsamplestraining,zhu2025llm}.

While a word-based approach to confusables can be effective~\cite{jia2020trainingwakeworddetection}, previous work has not attempted to exhaustively target all possible confusable sounds. Uncommon proper nouns and combinations of words outside the target language, if not present in the training set, may exhibit undefined behavior~\cite{191968}. Although such terms might be rare, they are still worth addressing~\cite{SCHONHERR2022101328}, as phrases that the KWS model is exposed to in productionized environments change over time as language evolves. If users discover alternative phrases, they may become adopted as undesired and unintended triggers~\cite{191968}.

One approach to identifying a large set of confusables by Gao et al.~\cite{gao2020dataefficientmodelingwakeword}
is to mine ASR transcripts for lexically similar words and phrases, and then treat the corresponding audio as a negative example. A related method is to search existing datasets for naturally occurring confusables~\cite{ai2024mmkwsmultimodalpromptsmultilingual}, however this requires additional data and may not be as comprehensive in covering the full spectrum of pronounciations as a systematic approach.  

By contrast, our approach uses TTS to ensure coverage of any lexically similar word or phrase, and generates large sets of confusables instead of mining them. While there are advantages to audio mining instead of synthesis, especially with respect to obtaining realistic audio, our approach does not require existing audio datasets and is not limited by the frequency of confusables in real audio nor by the quality of the ASR transcription of an uncommon word or phrase.

This paper addresses long-tail confusables by systematically and comprehensively mutating the keyword's graphemes. This involves making edits to the written form of the keyword with insertions, deletions, and substitutions. These edited keywords are used as negative examples to enhance KWS model robustness and accuracy. This approach leverages the fact that a small number of changes in spelling correspond to subtle changes in pronunciation, creating audio instances distinct from, but perceptually similar to, the true keyword. By exposing the KWS model to these carefully crafted confusables during training, we aim to improve its ability to discriminate between the keyword and its acoustic near-neighbors.

A close alternative to the grapheme edit approach is to edit phonetic sequences. Operating on phonemes more directly corresponds to audio differences, and would likely be more efficient and scale better to different languages. However scaling this approach would require both a grapheme-to-phoneme (G2P) model~\cite{Yolchuyeva_2019, qharabagh2024llmpoweredgraphemetophonemeconversionbenchmark} and a high-quality TTS engine for phonetic alphabets. We approach the problem using grapheme substitution in this paper to study the effects of edit mutations without being limited by the quality of these components.

In situations where real speech is not available, TTS systems have been increasingly used in KWS applications~\cite{9413448, lin2020trainingkeywordspotterslimited,park2024adversarialtrainingkeywordspotting}. It is particularly appropriate for our proposed method since the edited transcripts we generate are often words that a native speaker may find difficult to pronounce. Large numbers of confusables are also more time-consuming and costly to collect.

To improve the quality of our synthesized audio, we use style transfer, which is further described in Section \ref{ssec:audiolmtts}. Style transfer has been previously used for KWS~\cite{park24_syndata4genai}, but its contributions to KWS have not been well-quantified.

The major contributions of this paper include:

\begin{itemize}
    \item GraphemeAug, an algorithm to systematically generate confusables. To our knowledge, this is the first paper to explore systematic generation of synthetic confusables.
    \item We show the effectiveness of GraphemeAug confusables on a similar test set without degrading the ability to detect the target keyword.
    \item We show that when training a model with only TTS data, TTS with style transfer improves KWS model quality compared to TTS without style transfer.
    \item We find that the number of unique confusables used in training affects model quality, and recommend using as many as possible.
    \item We show that more mutations also improves model quality, but has a smaller effect.
    \item We report training on systematic confusables improves performance on natural confusable words and phrases, even when all confusables are generated by synthetic phrase mutations and TTS. The converse, of training on natural confusables and testing on systematic confusables, is less effective.
\end{itemize}

% The rest of this paper defines the algorithm to generate confusables, demonstrates its impact on negative datasets that are designed to be challenging, and outlines how the diversity of confusables makes a large impact on model performance.
\section{Methodology}
\label{sec:methodology}

Figure~\ref{fig:data-creation-figure} describes the training architecture.

\subsection{GraphemeAug algorithm}
\label{ssec:keyword_editing}

This paper introduces an algorithm for generating confusables based on the systematic modification of keyword graphemes. The algorithm operates by applying three core edit operations to the target keyword:

\begin{enumerate}
  \item \textbf{Grapheme Addition}: A single grapheme is inserted at a position within the keyword.
  \item \textbf{Grapheme Removal}: A single grapheme is removed from the keyword.
  \item \textbf{Grapheme Substitution}: A single grapheme is replaced with another grapheme of the same class (vowel or consonant).
\end{enumerate}

The edit distance is the Levenshtein distance between the confusable and target keyword. A combination of all 3 operations can be used to generate a confusable as seen in Table~\ref{tab:filter_words}. We implemented a recursive algorithm that edits the keyword to generate all possible confusables.

\begin{table}[H]
    \centering
    \small
    \caption{Examples of confusables created by editing the keyword ``Hey Google''. Because the algorithm disregards linguistic rules, outputs may not follow typical language patterns.}
    \vspace{-5pt}
    \begin{tabular}{c c c}
        \toprule
        \textbf{Edit Distance} & \textbf{Operations} &  \textbf{Confusable} \\  
        \midrule
        1 & G\textrightarrow P & Hey Poogle \\
        1 & Del Y & He Google \\
        2 & H\textrightarrow R, O\textrightarrow U & Rey Gougle \\
        % 2 & Ins V, Ins L & Hevy Gologle \\
        % 3 & H\textrightarrow R, O\textrightarrow U, E \textrightarrow A & Rey Gougla \\
        3 & Ins V, Ins L, E\textrightarrow U & Hevy Gologlu \\
        \bottomrule
    \end{tabular}
    \label{tab:filter_words}
    \vspace{-10pt}
\end{table}

% \begin{table}[H]
%     \centering
%     \small
%     \begin{tabular}{c c c}
%         \toprule
%         \textbf{Keyword} & \textbf{Edit Distance} & \textbf{\# Confusables} \\  
%         \midrule
%         Hey Google & 1 & 433 \\
%         Hey Google & 2 & 92,789 \\
%         Hey Google & 3 & 13,263,534 \\
%         \bottomrule
%     \end{tabular}
%     \caption{Number of confusables for each edit distance for the keyword ``Hey Google".}
%     \label{tab:filter_word_counts}
% \end{table}

% The set of words grows exponentially with increased edit distances, as shown in Table~\ref{tab:filter_word_counts}.

\subsection{AudioLM TTS}
\label{ssec:audiolmtts}

All the experiments described in this paper use TTS data for training. We use an AudioLM-based model~\cite{park24_syndata4genai, borsos2023audiolmlanguagemodelingapproach, kharitonov2023speakreadprompthighfidelity} which has two modes, one with style transfer and the other that randomly samples a completely synthetic speaker. When style transfer is used, this model is able to mimic the voice of the input audio. Specifically, it patterns the synthesized audio after the prosody and speaker characteristics of the original example. This ensures that the synthesized examples retain some of the diversity and naturalness present in the original source data. For simplicity and consistency, real data was used only for generating TTS examples but not for training, although it would likely improve the quality of a model intended for production use. 
\begin{figure*}[ht]
\centering
\begin{subfigure}[t]{0.32\textwidth}
\includegraphics[width=1\columnwidth]{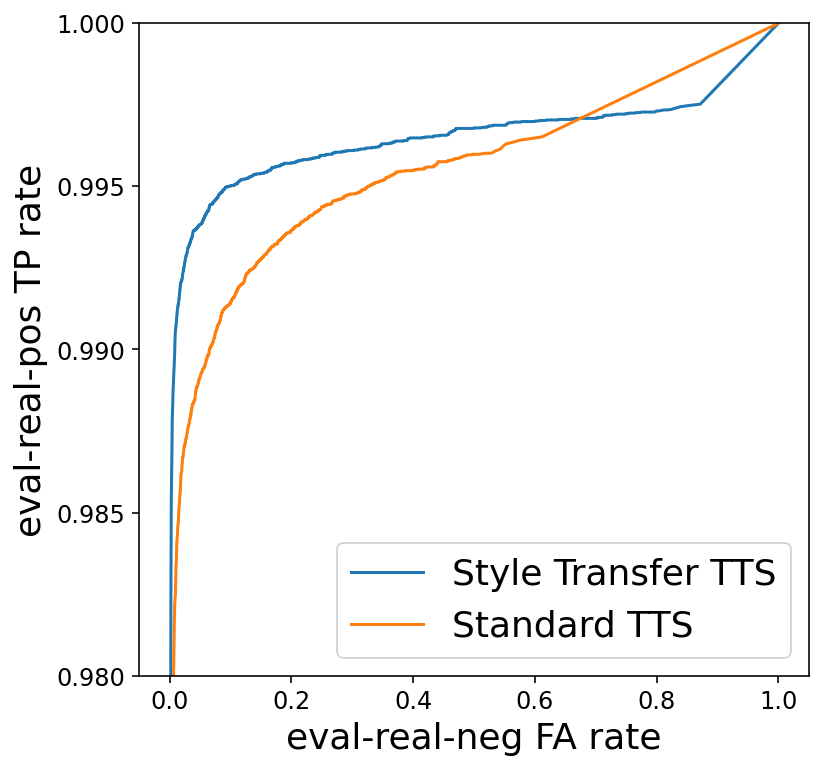}
\caption{}
\label{fig:personalized-roc}
\end{subfigure}
\begin{subfigure}[t]{0.31\textwidth}
\includegraphics[width=1\columnwidth]{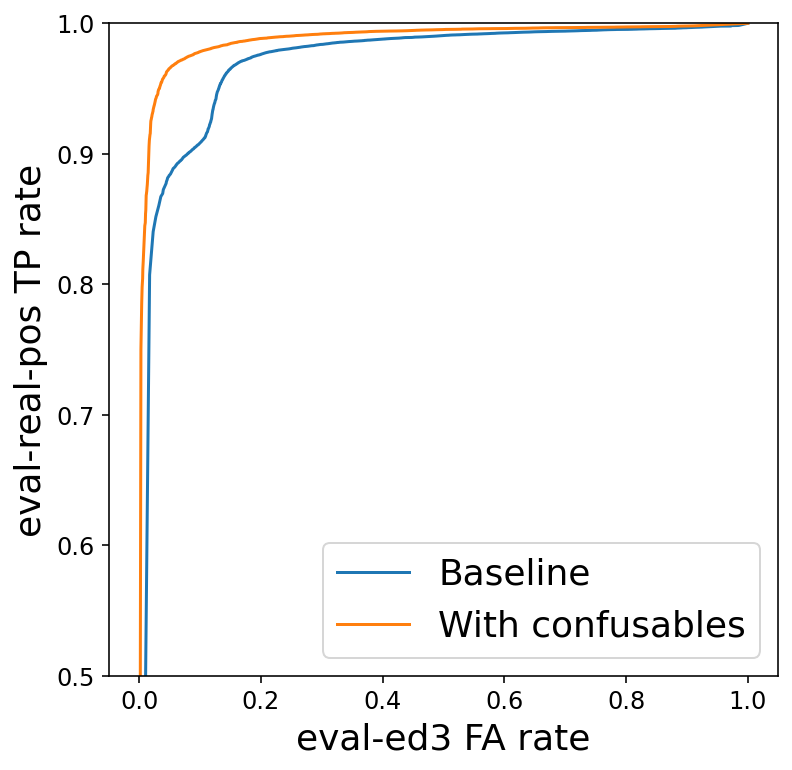}
\caption{}
\label{fig:ed3-10k-improves-on-ed3}
\end{subfigure}
\begin{subfigure}[t]{0.32\textwidth}
\includegraphics[width=1\columnwidth]{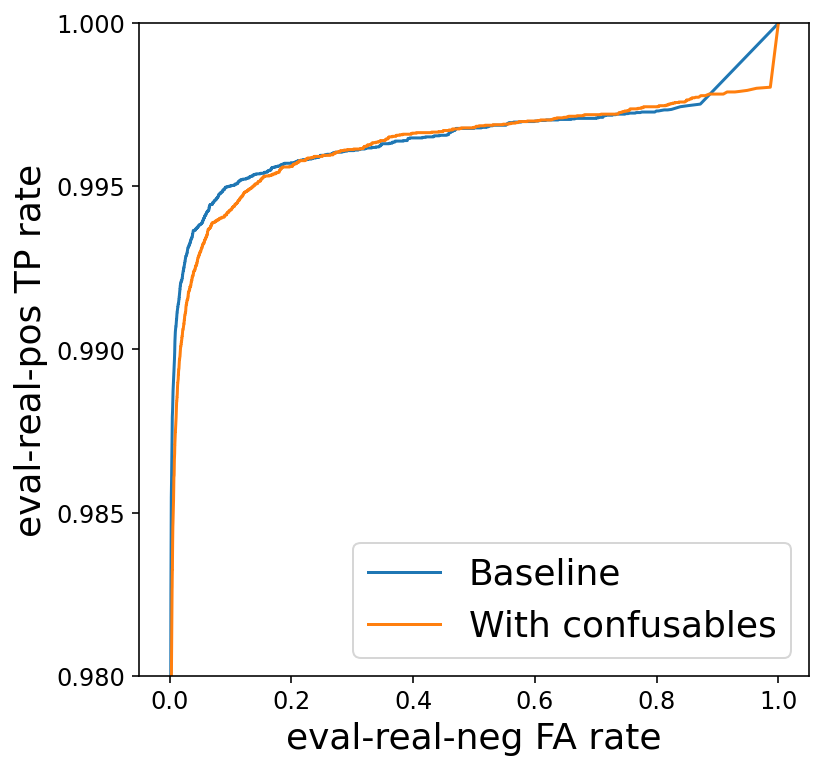}
\caption{}
\label{fig:ed3-10k-stays-same}
\end{subfigure}
\vspace{-10pt}
\caption{ROC curves. (a) shows that TTS with style transfer generally outperforms standard TTS. (b) shows how adding confusables improves performance when comparing real positive examples to negatives synthesized for the same confusables, but spoken by other users. (c) shows that performance of the model with confusables shows similar quality on real positive and real non-confusable negative data.}
\vspace{-10pt}
\end{figure*}

\section{Model and data}
\label{sec:model}

\subsection{Data}
\label{ssec:dadta}

\subsubsection{Baseline data}
\label{ssec:trainingdata}

The data used for training was generated from the TTS model described in Section~\ref{ssec:audiolmtts}. We utilized 13 source datasets containing real human speech from various English locales, each with approximately 600,000 examples. These source datasets contain a placeholder keyword and a query (e.g., ``Hey Indy, what's the weather?''). 

To create positive examples, we replaced the placeholder keyword with the target keyword (``Hey Google'') in the transcript of each source audio example. This modified transcript, along with the original audio, was then passed to the TTS system. To create non-confusable negative examples, we deleted the placeholder keyword from the transcript (e.g., ``Hey Indy what's the weather'' replaced with ``what's the weather'').

The synthesized datasets were transformed into 25 variations by applying room simulation and noise mixing~\cite{7178863,kim17_interspeech}. Finally, filterbank energies were calculated over a 25ms window at every 10ms time frame. Three consecutive frames from the 40-dimensional vector were stacked to create the 120-dimensional input feature vector, as prepared in Raziel and Park~\cite{raziel2019endtoendstreamingkeywordspotting}. The total size of datasets were 195 million positive examples and 195 million negative examples. Baselines were trained on an equal ratio of positives and negatives generated in this way.

Two versions of positive and negative datasets were generated, one with style transfer and the other using standard TTS.

% \subsubsection{Baseline data without style transfer}
% \label{ssec:baseline-data-no-style-transfer}

\subsubsection{Confusion training data}
\label{ssec:confusiondata}

The datasets containing confusables were generated using the same pipeline as the baseline data. The same source datasets were used, except that the target keyword was replaced with a confusable phrase. We generated datasets with various edit distances (1, 2, 3) and counts of unique confusables (10, 100, 1k, 10k) to explore the impact of confusable diversity on model performance. When training on confusable datasets, we replaced 10\% of the negative examples with confusables (Figure~\ref{fig:data-creation-figure}).

% \begin{table}[H]
%     \centering
%     \small
%     \begin{tabular}{c c c}
%         \toprule
%         \textbf{Dataset Name} & \textbf{Edit Distance} & \textbf{\# Confusables} \\  
%         \midrule
%         train-pos & - & - \\
%         train-neg & - & - \\
%         % 1 & 10 & ed1-confs10 \\
%         train-ed1-100 & 1 & 100 \\
%         % 1 & 433 & ed1-confs1k \\
%         % 2 & 10 & ed2-confs10 \\
%         train-ed2-100 & 2 & 100  \\
%         % 2 & 1000 & ed2-confs1k \\
%         % 2 & 10000 & ed2-confs10k \\
%         train-ed3-10 & 3 & 10 \\
%         train-ed3-100 & 3 & 100 \\
%         train-ed3-1k & 3 & 1000  \\
%         train-ed3-10k & 3 & 10000  \\
%         \bottomrule
%     \end{tabular}
%     \caption{The various datasets: positive, plain negative and confusable negative datasets. This shows the number of unique confusables. The number of total utterances in every dataset is 180 million with approximately 88355 hours of audio.}
%     \label{tab:training_set_names}
% \end{table}

% The threshold for the model is selected by fixing the false accept rate to one per hour on the eval-real-neg dataset containing around 20,190 random phrases (not confusables) from en-US speakers. 

\subsubsection{Evaluation data}
\label{ssec:evaluationdata}

A positive and negative dataset were crowdsourced, and contain real spoken speech. The positive dataset consists of 61736 examples and the negative dataset consists of 20190 examples. We label these ``eval-real-pos'' and ``eval-real-neg''. To evaluate effectiveness in more realistic conditions, we also crowdsourced a dataset of 3779 confusable examples containing English words and phrases. The set of confusables was handpicked to represent likely phrases that users might say, reflecting natural language patterns. This dataset is labeled ``eval-real-conf''.

Synthetic evaluation data was also generated using the same pipeline as the training data, using a withheld set of source audio for source speaker and prosody matching. We built a dataset of 9595 synthetic examples with randomly sampled confusables of edit distance 3. This dataset is labelled ``eval-ed3''.

% We created a positive evaluation dataset and negative confusable datasets of different edit distances. These confusable datasets contain up to 10,000 unique confusables unless restricted by the possible unique confusables due to the edit distance being small. In some cases, mutated confusables are distinct in their spelling but are identical phonetically. We removed any example whose ASR transcript contained the hotword as the top hypothesis.

% \begin{table}[H]
%     \centering
%     \small
%     \begin{tabular}{c c c c c}
%         \toprule
%         \textbf{Name} & \textbf{Type} & \textbf{\# Utts} & \textbf{Edit Distance} & \textbf{Audio}  \\  
%         \midrule
%         eval-real-pos & Pos & 9660 & - & TTS \\
%         eval-ed3 & Neg & 9595 & 3 & TTS  \\
%         eval-real-conf & Neg & 3779 & - & Real \\
%         eval-real-neg & Neg & 20190 & - & Real \\
%         \bottomrule
%     \end{tabular}
%     \caption{The evaluation datasets used to measure performance.}
%     \label{tab:test_set_name}
% \end{table}

\subsection{Model architecture}
\label{ssec:architechture}

The architecture employed in this study utilized a two-stage encoder-decoder structure optimized for streaming inference. The model consists of seven factored convolution layers (SVDF) and three bottleneck projection layers, totaling approximately 320,000 parameters. The architecture resembles~\cite{park24_syndata4genai, 6854370, raziel2019endtoendstreamingkeywordspotting}.

% \subsection{Experimental setup}
% \label{ssec:experimentalsetup}

% Our experimental setup uses the target keyword ``Hey Google'', which has the structure of a typical keyword used in assistant systems. We weight the number of positive and negatives equally in the baseline. When adding confusable datasets, we replace 10\% of the negative examples with confusables (Figure~\ref{fig:data-creation-figure}).

\begin{table*}[t]
    \centering
    % \small
    \begin{tabular}{c c c c}
        \toprule
        \textbf{Training Set} & \textbf{eval-real-neg} &  \textbf{eval-ed3} & \textbf{eval-real-conf} \\  
        \midrule
        Baseline (no style transfer) & 99.52 & 95.7 & 97.8 \\
        Baseline & 99.65 & 96.9 & 97.8 \\
        Baseline + TTS confusables from eval-real-conf & 99.57 & 91.7 & 99.0 \\
        Baseline + TTS confusables from GraphemeAug with edit distance 1 & 99.61 & 98.2 & 99.0 \\
        Baseline + TTS confusables from GraphemeAug with edit distance 3 & 99.63 & 98.8 & 98.9 \\
        \bottomrule
    \end{tabular}
    \caption{AUC (\%) from the ROC curve of eval-real-pos against various negative datasets. Unless otherwise stated, all datasets use TTS with style transfer. Models trained with confusables from GraphemeAug improve on targeted and untargeted confusables while maintaining performance on general negative evaluation sets.}
    \label{tab:overall-table-results}
    \vspace{-10pt}
\end{table*}

\section{Experimental Results}
\label{sec:experiments}

Training continued until loss curves stabilized, around 400,000 steps. We let training run to 800,000 steps and collected 10 checkpoints roughly equally spaced within that range. These checkpoints showed substantial noise in their AUC. The results below are the mean of the 10 AUC values. When plotting ROC curves, we plotted the curve with the median AUC.

\subsection{Style Transfer TTS improves quality}
\label{ssec:personalized TTS}

We see in Figure~\ref{fig:personalized-roc} that TTS with style transfer outperforms standard TTS at most False Accept (FA) rates. By using TTS with style transfer, the AUC improves by 22\%.
% Commenting out the statistical significance tests, as we don't run tests everywhere we could, and they probably open up more lines of criticism than they
% shut down.
%, this improvement is verified to be statistically significant using an ANOVA test on the AUC of 10 checkpoints after convergence with a p-value of 2e-6.
Because of this finding, the rest of this paper's results and figures use TTS with style transfer for its quality improvements.

\subsection{Introducing confusables improves quality}
\label{ssec:first-introducing-confusables}

We ran an experiment comparing a baseline trained with only standard positive and negative examples against a model that is trained with confusables of edit distance 3. We incorporated a dataset that sampled from 10,000 unique confusables, and replaced 10\% of the negative examples with such confusables. 

Figure~\ref{fig:ed3-10k-improves-on-ed3} shows that introducing confusables can substantially improve performance on confusable datasets. By training with confusables, the AUC improves by 61\%. Often there are tradeoffs in False Reject Rates (FRR) and False Accept Rates~\cite{raju2018dataaugmentationrobustkeyword,article}, however introducing GraphemeAug confusables with edit distance 3 did not degrade the performance on our positive dataset as shown in Figure~\ref{fig:ed3-10k-stays-same} and Table~\ref{tab:overall-table-results}.
% An ANOVA test (using the same method as described in Section \ref{ssec:personalized TTS}) did not detect a statistically significant difference.

\subsection{Number of confusables and edit distance}
\label{ssec:number-of-confusables}

\begin{figure}[ht]
\centering
\includegraphics[width=.8\columnwidth]{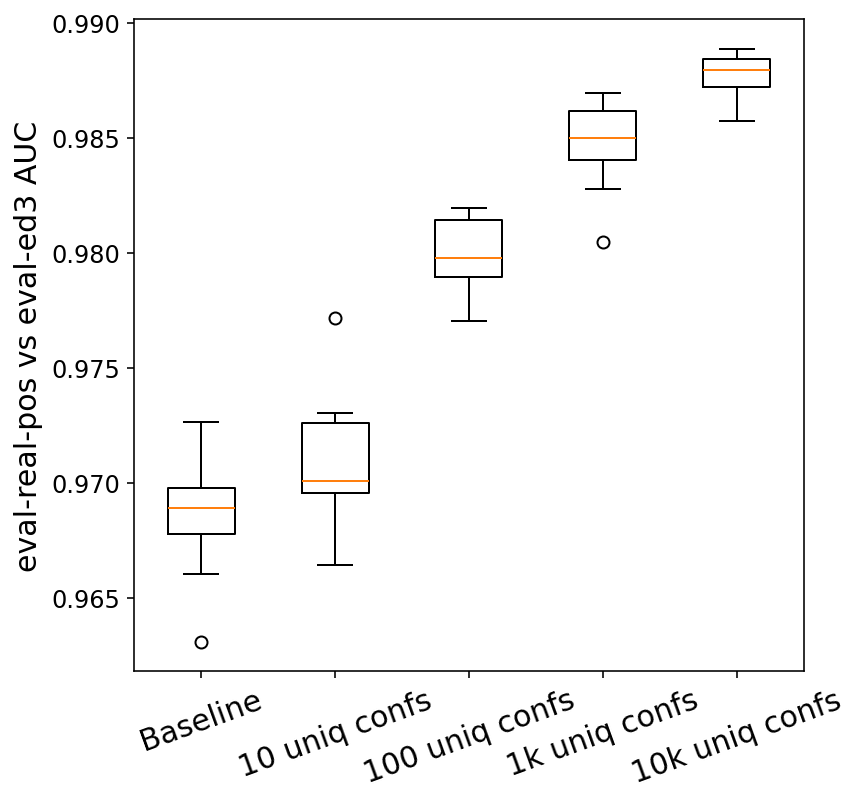}
\vspace{-5pt}
\caption{eval-real-pos vs eval-ed3 AUC when training with different number of unique confusables with edit distance 3.}
\label{fig:number-of-confusables}
\vspace{-10pt}
\end{figure}

We further conducted ablation studies on how the number of confusables and edit distance impact performance. To determine whether the model is able to generalize well and reject unseen confusables, we ablated the number of unique confusables. Figure~\ref{fig:number-of-confusables} shows that AUC improves as more confusables are introduced, increasing 58\% when trained with 10,000 confusables vs only 10. This suggests that the model's ability to generalize after seeing a small number of confusables is not assured, and that the model benefits from training on a diverse set of phrases~\cite{liu2023selectiontexttospeechdataaugment, zhu2024synth4kwssynthesizedspeechuser}. This highlights the importance of scale, which systematic methods like GraphemeAug can provide.

\begin{figure}[ht]
\centering
\includegraphics[width=.8\columnwidth]{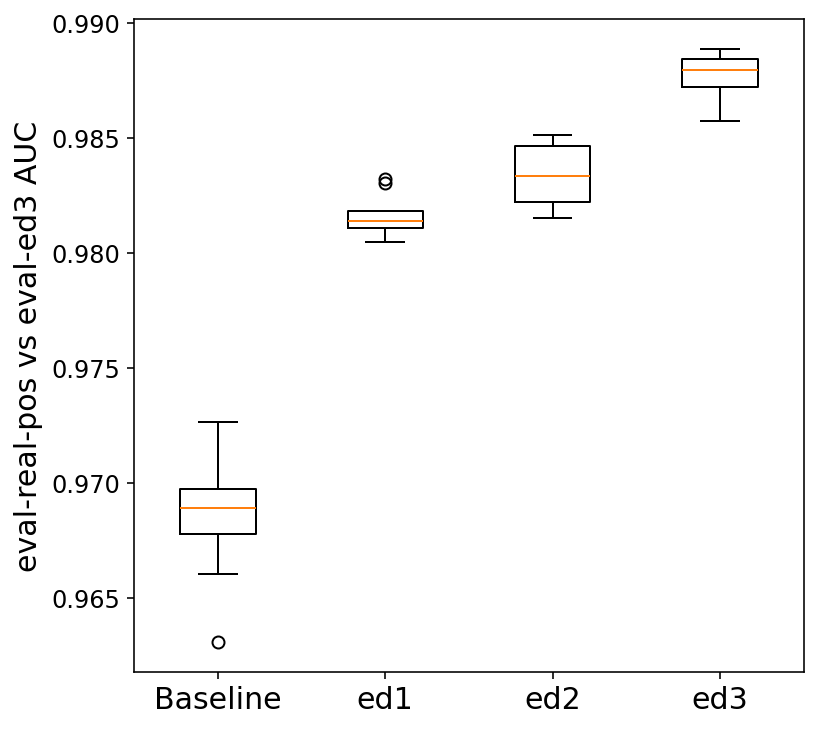}
\vspace{-5pt}
\caption{eval-real-pos vs eval-ed3 AUC when training with 10,000 unique confusables from different edit distances.}
\label{fig:various-edit-distances}
\vspace{-10pt}
\end{figure}

The number of edits applied to the keyphrase also has an effect, but a weaker one than the number of unique confusables used in training (see Figure~\ref{fig:various-edit-distances}).
Edit distances that are too low may lead to datasets that have many examples that sound almost identical to the keyword, leading to more false rejects. Setting the edit distance to be sufficiently large may help generate similar yet phonetically distinct phrases.

\subsection{GraphemeAug on real confusables}
\label{ssec:real-confusables}

To explore the applicability of GraphemeAug on unseen confusables and confusables that contain real spoken audio, we also evaluated on the ``eval-real-conf'' dataset.
For this analysis, we trained a model on confusables with only one edit. This model improves ``eval-real-conf'' AUC by 54\% compared to the baseline, showing that training on synthetic confusables can be effective at suppressing false accepts from real audio recordings of confusables.

We also explored whether the reverse is true---whether a model trained on real confusables is effective at suppressing synthetic confusables. We trained a model on TTS data generated from the ``eval-real-conf'' transcripts using a similar process to Figure~\ref{fig:data-creation-figure}. Table~\ref{tab:real-confusables} shows that while a model trained with synthetic confusables reaches a high AUC when tested both on synthetic confusables and real confusables, a model trained with real confusables reaches a high AUC only when tested on real confusables. When tested on synthetic confusables, it does much worse (91.7). This happens despite the model trained with real confusables having access to many more unique confusable training examples (3779) than the model trained on a single-edit confusables (433). This result suggests that training on synthetic confusables provides benefits that real confusable datasets do not easily provide.

% and then tested both this model and the model with confusables with a single edit

% models against test sets containing the same kind of confusables, and the opposite kind, as shown in Table~\ref{tab:real-confusables}.

% compare to the real confusable dataset's 3779 examples.

% Despite this advantage, the model trained on real confusables performed substantially worse, with an AUC of 0.083 compared to AUCs around 0.01 for all other conditions.

% we compare against a model trained specifically on the same confusables as those in the evaluation dataset. While we are not able to achieve an identical AUC on the eval-real-conf dataset, we have only a 17\% delta. However the opposite is not true. When evaluating on the eval-ed3 dataset, a model trained on GraphemeAug confusables produces an AUC 85\% lower than one trained on hand selected confusables. This suggests that the systematically generated confusables are largely able to cover a wide range of confusable pronounciations. However, a carefully selected set of limited confusables is unable to cover the range of systematically generated confusables, meaning it may miss other unseen phrases which may become false triggers.

\begin{table}[H]
    \centering
    \small
    \caption{Mean AUC (\%) comparing training on TTS data with confusables from the eval-real-conf dataset and confusables with edit distance 1.}
    \vspace{-5pt}
    \begin{tabular}{c c c c}
        \toprule
        \textbf{Training Set} & \textbf{eval-ed3} & \textbf{eval-real-conf}  \\  
        \midrule
        train-ed1 & 98.2 & 99.01 \\
        train-real-conf & 91.7 & 99.04 \\
        \bottomrule
    \end{tabular}
    
    \label{tab:real-confusables}
    \vspace{-10pt}
\end{table}

\section{Conclusion}
\label{sec:conclusion}

This paper presents a novel approach to improve keyword spotting (KWS) system robustness by systematically generating confusables based on grapheme mutations of the target keyword.  Our method creates a diverse set of acoustically similar phrases, mimicking both common and rare pronunciation variations. We demonstrate that applying style transfer to TTS training data improves model quality, that synthetic confusables improve model quality on similar confusables without degrading general KWS performance, that the number of unique confusables influences model performance (Figure~\ref{fig:number-of-confusables}), that the number of mutations has a smaller effect, and that synthetic confusables help performance on real audio confusables more than training on real audio confusables helps performance on synthetic confusables.

% in conclusion, our experiments, using ``hey google'' as the target keyword, demonstrate the effectiveness of our approach, showing a significant reduction in false accept rates on both real (70\%) and synthetic (83\%) confusion datasets.

% our grapheme mutation method offers a powerful and efficient way to generate confusion phrases for training robust KWS systems.

\clearpage
\ifinterspeechfinal
    \section{Acknowledgements}
\label{sec:ack}

We thank Cheng-Chieh Peng, Justin Chen and Rentao Sun for their collaboration in this work.
\else
\fi

\bibliographystyle{IEEEtran}
\bibliography{mybib}

\end{document}